\documentclass[11pt,dvips]{article}
\usepackage{eurosym}
\usepackage{graphicx}
\usepackage{amsmath}
\usepackage{amssymb}
\usepackage{latexsym}
\usepackage{color}
\usepackage{epstopdf}

\input{tcilatex}
\begin{document}

\thispagestyle{empty}

\begin{center}
\vspace{1.8cm}

{\Large \textbf{Quantifying Quantumness of Correlations Using Gaussian R\'{e}%
nyi-2 Entropy in Optomechanical Interfaces}}

\vspace{1.5cm}

\textbf{Jamal El Qars}$^{a}${\footnote{%
email: \textsf{j.elqars@gmail.com}}}, \textbf{Mohammed Daoud}$^{b,c}${\footnote{%
email: \textsf{m$_{-}$daoud@hotmail.com}}} and \textbf{Rachid Ahl Laamara}$%
^{a,d} ${\footnote{%
email: \textsf{ahllaamara@gmail.com}}}

\vspace{0.5cm}

$^{a}$\textit{LPHE-MS, Faculty of Sciences, Mohammed V University of Rabat,
Rabat, Morocco}\\[0.5em]
$^{b}$\textit{Abdus Salam International Centre for Theoretical Physics,
Miramare, Trieste, Italy}\\[0.5em]
$^{c}$\textit{Department of Physics, Faculty of Sciences, University Hassan
II, Casablanca, Morocco}\\[0.5em]
$^{d}$\textit{Centre of Physics and Mathematics (CPM), Mohammed V University
of Rabat, Rabat, Morocco}\\[0.5em]

\vspace{3cm} \textbf{Abstract}
\end{center}

\baselineskip=18pt Using the Gaussian R\'{e}nyi-2 entropy, we analyze the
behavior of two different aspects of quantum correlations (entanglement and
quantum discord) in two optomechanical subsystems (optical and mechanical).
We work in the resolved sideband and weak coupling regimes. In
experimentally accessible parameters, we show that it is possible to create
entanglement and quantum discord in the considered subsystems by quantum
fluctuations transfer from either light to light or light to matter. We find
that both mechanical and optical entanglement are strongly sensitive to
thermal noises. In particular, we find that the mechanical one is more
affected by thermal effects than that optical. Finally, we reveal that under
thermal noises, the discord associated with the entangled state decays
aggressively, whereas the discord of the separable state (quantumness of
correlations) exhibits a freezing behavior, seeming to be captured over a
wide range of temperature. 
\newpage

\section{Introduction}

It is well believed that entanglement is a very important ingredient for
quantum information processing and quantum communication tasks that cannot
be done efficiently classically \cite{Nielsen}. It plays a central role in
several tasks such as, quantum teleportation, quantum key distribution, and
superdense coding \cite{Loock,HorodeckiS}. Previously, any correlation
without entanglement was thought to be purely classical as they can be
produced with local operations and classical communications (LOCC) \cite%
{Kavan Modi}. However, it has been shown in many theoretical \cite%
{Braunstein-} and experimental \cite{Lanyon} works that
separable(unentangled) states, traditionally referred to as \textit{%
classically correlated,} might retain some signatures of quantumness with
potential operational applications for quantum tasks \cite{Kavan Modi,G(1)}.
For example, quantum correlations without entanglement were shown to be
useful to characterize resources in \cite{RahimiKeshari}: a quantum
computational model (DQC1); quantum state merging; remote state preparation;
encoding information onto a quantum state; quantum phase estimation; and
quantum key distribution. In general, entanglement is difficult to generate,
more fragile and rapidly decays to zero in dissipative-noisy systems \cite%
{ESD}. Conversely, it has been pointed out that quantum correlations without
entanglement are robust, exhibiting some peculiar features against Markovian
decoherence \cite{Kavan Modi}. We can cite for instance \cite{NSP}: they can
never suffer the sudden death; within a dissipative-noisy systems, they can
reach an asymptotic nonzero value (i.e., freezing behavior) even for high
temperatures; they can exhibit a sudden change from the \textit{classical
decoherence} regime to the \textit{quantum decoherence} regime. Such
interesting properties make, therefore, this kind of quantum correlations
more desirable resource in quantum information processing, opening new
avenues for theoretical exploration and practical applications \cite{Kavan
Modi}.

Having a \textit{quantum separable state}, the question is then how to
quantify its quantum correlations ? In fact, quantifying quantum
correlations of separable states,i.e.,\textit{the quantumness of correlations%
}, is one of the crucial tasks related to the understanding and efficient
exploitation of such states in different quantum information processing
schemes \cite{Kavan Modi}. In this sense, the quantification of the
quantumness of correlations, was carried out originally by means of the
quantum discord \cite{ZA}, defined and evaluated mainly for discrete
variables systems \cite{ZA} and subsequently extended to the continuous
variable systems \cite{G(1)}. Recently, a number of other related
quantifiers of the \textit{quantumness} of correlations were proposed. As
examples, one may cite the relative entropy of quantumness \cite{ERQ}, the
quantum deficit \cite{Marcio}, the negativity of quantumness \cite{Takafumi}%
, the Gaussian quantum discord \cite{G(1)}, the Gaussian geometric discord
\cite{AG}, the operational Gaussian Discord \cite{Rahimi}, the Gaussian
Hellinger distance \cite{Marian}, and also the Gaussian R\'{e}nyi-2 discord
\cite{AGS}.

In a double-cavity optomechanical system fed by broadband two-mode squeezed
light, we propose in this paper to investigate both entanglement and quantum
discord between two mechanical modes (the mechanical subsystem) on one hand,
and between two optical modes (the optical subsystem) on the other hand. To
accomplish this, we shall use respectively the Gaussian R\'{e}nyi-2
entanglement and the Gaussian R\'{e}nyi-2 discord to quantify entanglement
and quantum discord \cite{AGS}.

During the last few years, optomechanical devices based on the interaction
between light and mechanical degrees of freedom, have sparkled the interest
of a vast scientific community as very promising platform for demonstrating
various quantum phenomena \cite{Meystre}. Proposals include decoherence \cite%
{Ulrich}, quantum back-action evading \cite{Korppi}, macroscopic quantum
superposition \cite{MAbdi}, optomechanically induced transparency \cite%
{GAgarwal,OIT}, the cooling of an harmonic oscillator near its quantum
ground state \cite{Teufel,Chan,Yu}, quantum states transfer \cite%
{Rakhubovsky}, and also the study of various aspects of quantum
correlations: quantum entanglement \cite{entanglement, OET,MET}, quantum
discord \cite{discord} and quantum steering \cite{steering}.

The remainder of this paper is organized as follows. In Sec \ref{sec2}, we
introduce the optomechanical system at hand and we exactly solve the quantum
Langevin equations in Fourier space, which allows us to obtain analytical
formulas for the covariance matrices fully describing the subsystems under
investigation. Sec \ref{sec3} is devoted to the quantification of
nonclassical correlations in the two studied subsystems. This quantification
is performed using the Gaussian R\'{e}nyi-2 entropy. Finally, in Sec \ref%
{sec4} we draw our conclusions.

\section{ The optomechanical model and Hamiltonian \label{sec2}}

\subsection{ The model}

\begin{figure}[tbh]
\centerline{\includegraphics[width=14cm]{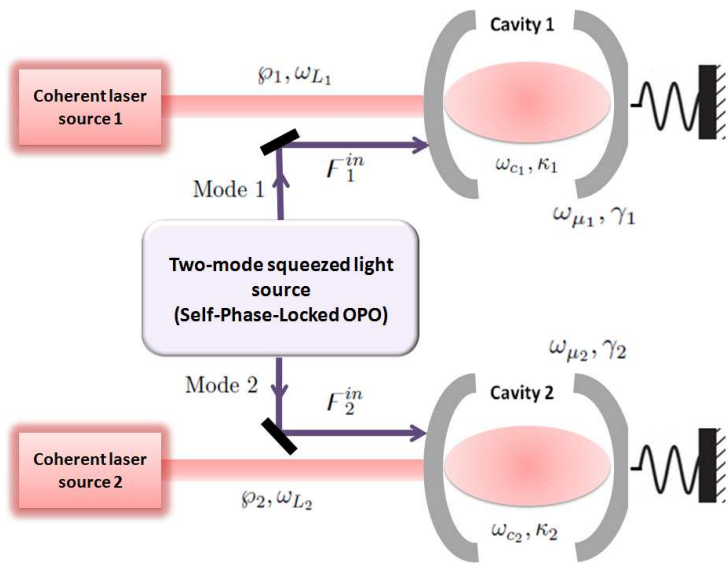}}
\caption{Sketch of two optomechanical Fabry-P\'{e}rot cavities fed by
non-classical light from \textit{self-phase-locked OPO}. Each cavity is
pumped by a coherent laser field of power ${\wp }_{j}$ and frequency $%
\protect\omega _{L_{j}}$ ($j=1,2$). The $j^{th}$ movable mirror will be
treated as an harmonic oscillator with frequency $\protect\omega _{\protect%
\mu _{j}}$, a damping rate $\protect\gamma _{j}$ and an effective mass $m_{%
\protect\mu _{j}}$. $\digamma _{j}^{in}$ is the $j^{th}$ noise operator
corresponding to the $j^{th}$ squeezed mode.}
\label{Fig.1}
\end{figure}
\noindent We consider two optomechanical Fabry-P\'{e}rot cavities (see Fig. %
\ref{Fig.1}). Each cavity having length $l_{j}$ and frequency $\omega
_{c_{j}}$, is constituted by two mirrors. The first one is fixed and
partially transmitting. Whereas the second is a totally reflecting movable
mirror. This end is modeled as single-mode quantum harmonic oscillator with
effective mass $m_{\mu _{j}}$, oscillating with the frequency $\omega _{\mu
_{j}}$ and damped with a rate $\gamma _{j}$ \cite{mmm}. In addition, the $%
j^{th}$ cavity is pumped by a laser field with the input power ${\wp }_{j}$,
phase $\varphi _{j}$ and frequency $\omega _{L_{j}}$ ($j=1,2$). Finally, the
two cavities are jointly pumped by a two-mode squeezed light produced for
example by a self-phase-locked optical parametric operator (OPO) below
threshold \cite{Keller}.

\subsection{ The Hamiltonian}

The Hamiltonian of the two cavities, in a frame rotating at the frequency of
the lasers, reads \cite{Law}
\begin{equation}
H=\sum_{j=1}^{2}\left[ \hbar\left( \omega _{c_{j}}-\omega _{L_{j}}\right)
c_{j}^{\dag }c_{j}+\hbar\omega _{\mu _{j}}b_{j}^{\dag }b_{j}+\hbar
g_{j}c_{j}^{\dag }c_{j}(b_{j}^{\dag }+b_{j})+\hbar\varepsilon
_{j}(e^{i\varphi _{j}}c_{j}^{\dag }+e^{-i\varphi _{j}}c_{j})\right],
\label{E1}
\end{equation}%
where $b_{j}$($c_{j}$) is the annihilation operator of the $j^{th}$
mechanical(optical) mode. Moreover, $g_{j}=\left( \omega
_{c_{j}}/l_{j}\right) \sqrt{\hbar /m_{\mu _{j}}\omega _{\mu _{j}}}$ is the
vacuum optomechanical coupling rate between single photon and single phonon.
The parameter $\varepsilon _{j}$ is related to the input laser power ${\wp }%
_{j}$ by $\varepsilon _{j}=\sqrt{2\kappa _{j}{\wp }_{j}/\hbar \omega _{L_{j}}%
}$, where $\kappa _{j}$ is the photon decay rate by leaking out of the $%
j^{th}$ cavity.

\subsection{Quantum Langevin equation}

A proper analysis of the system at hand must include the dissipations and
the noises in the cavities. This can be accomplished by considering the
following set of nonlinear quantum Langevin equations (QLEs)
\begin{eqnarray}
\dot{b}_{j} &=&-(\gamma _{j}/2+i\omega _{\mu _{j}})b_{j}-ig_{j}c_{j}^{\dag
}c_{j}+\sqrt{\gamma _{j}}\zeta _{j}^{in},  \label{E2} \\
\dot{c}_{j} &=&-\left( \kappa _{j}/2-i\Delta _{j}\right)
c_{j}-ig_{j}c_{j}(b_{j}^{\dag }+b_{j})-i\varepsilon _{j}e^{i\varphi _{j}}+%
\sqrt{\kappa _{j}}\digamma _{j}^{in},  \label{E3}
\end{eqnarray}%
where $\Delta _{j}=\omega _{L_{j}}-\omega _{c_{j}}$ is the $j^{th}$ detuning
\cite{mmm}. Moreover, $\zeta _{j}^{in}$ is the $j^{th}$ Brownian noise
operator with zero mean value ($\langle \zeta _{j}^{in}\rangle =0$),
describing the coupling of the $j^{th}$ movable mirror with its own
environment. In general, $\zeta _{j}^{in}$ is not a Markovian Gaussian noise
(not $\delta $-correlated). However, for large mechanical quality factors $%
\mathcal{Q}_{j}=\omega _{\mu _{j}}/\gamma _{j}\gg 1$, which is a necessary
condition to achieve quantum effects \cite{Vitali(1)}, the operator $\zeta
_{j}^{in}$ becomes $\delta $-correlated \cite{Girish}
\begin{equation}
\left( \langle \zeta _{j}^{in\dag }(\omega )\zeta _{j}^{in}(\omega ^{\prime
})\rangle ;\langle \zeta _{j}^{in}(\omega )\zeta _{j}^{in\dag }(\omega
^{\prime })\rangle \right) =2\pi \Big(n_{th,j};(n_{th,j}+1)\Big)\delta
(\omega +\omega ^{\prime }),  \label{E4-}
\end{equation}%
with $n_{th\mathrm{,}j}=1/\left( \exp (\hbar \omega _{\mu
_{j}}/k_{B}T_{j})-1\right) $ is the $j^{th}$ mean thermal photons number at
the mechanical frequency $\omega _{\mu _{j}}$. $T_{j}$ and $k_{B}$ are
respectively the temperature of the environment surrounding the $j^{th}$
movable mirror and the Boltzmann constant. Another kind of noise affecting
the $j^{th}$ cavity is the input squeezed vacuum noise operator $\digamma
_{j}^{in}$, with zero mean value ($\langle \digamma _{j}^{in}\rangle =0$). In what follows, we assume that the squeezed field is resonant with the
cavities, which is required in order to achieve optimal quantum states transfer from
squeezed light to either mechanical degrees of freedom or optical modes \cite{Hammerer}. The operator $\digamma _{j}^{in}$ has the following nonzero frequency-domain correlation functions \cite{JieLi}
\begin{eqnarray}
\left( \langle \delta \digamma _{j}^{in^{\dag }}(\omega )\delta \digamma
_{j}^{in}(\omega ^{\prime })\rangle ;\langle \delta \digamma
_{j}^{in}(\omega )\delta \digamma _{j}^{in^{\dag }}(\omega ^{\prime
})\rangle \right)  &=&2\pi \Big(N;(N+1)\Big)\delta (\omega +\omega ^{\prime
})\text{ ,}  \label{E7} \\
\left( \langle \delta \digamma _{j}^{in}(\omega )\delta \digamma
_{k}^{in}(\omega ^{\prime })\rangle ;\langle \delta \digamma _{j}^{in^{\dag
}}(\omega )\delta \digamma _{k}^{in^{\dag }}(\omega ^{\prime })\rangle
\right)  &=&2\pi \Big(M;M\Big)\delta (\omega +\omega ^{\prime }-\omega _{\mu
_{1}}-\omega _{\mu _{2}})\text{ },  \label{E9}
\end{eqnarray}%
for $j\neq k=1,2$ with $N=\mathrm{sinh}^{\mathrm{2}}r$ and $M=\mathrm{sinh}r%
\mathrm{cosh}r$, $r$ being the squeezing parameter.

\subsection{Linearization of quantum Langevin equations}

To study quantum correlations in the mechanical and optical subsystems under
investigation, we need to calculate their fluctuations. Assuming that the
nonlinear coupling between the $j^{th}$ cavity mode and its corresponding
mechanical mode is weak \cite{Girish}, the fluctuation $\delta c_{j}$($%
\delta b_{j}$) of the operator $c_{j} $($b_{j}$) is much smaller than its
steady-state mean value $\langle c_{j}\rangle $($\langle b_{j}\rangle $).
Thus we can linearize the system around the steady-states by writing the
operator $c_{j}$($b_{j}$) as the sum of its steady-state value $\langle
c_{j}\rangle $($\langle b_{j}\rangle $) and a small linear displacement $%
\delta c_{j}$ ($\delta b_{j} $), i.e., $c_{j}=\langle c_{j}\rangle +\delta
c_{j}$ and $b_{j}=\langle b_{j}\rangle +\delta b_{j}$. The mean values $%
\langle c_{j}\rangle \equiv c_{js}$ and $\langle b_{j}\rangle \equiv b_{js}$
can be easily calculated by setting the time derivatives to zero and
factorizing the averages in Eqs. (\ref{E2})-(\ref{E3}). One obtains
therefore
\begin{equation}
\langle c_{j}\rangle =c_{js}=\frac{-i\varepsilon _{j}e^{i\varphi _{j}}}{%
\kappa _{j}/2-i\Delta _{j}^{\prime }}\text{ \ \ \ and \ \ \ }\langle
b_{j}\rangle =b_{js}=\frac{-ig_{j}\left\vert c_{js}\right\vert ^{2}}{\gamma
_{j}/2+i\omega _{\mu _{j}}},\qquad  \label{E10}
\end{equation}%
with $\Delta _{j}^{\prime }$ $=$ $\Delta _{j}$ $-g_{j}(b_{js}^{\ast
}+b_{js}) $ is the $j^{th}$ effective cavity detuning including the
radiation pressure effects \cite{mmm}. We suppose henceforth that the two
cavities are strongly driven $\left\vert c_{js}\right\vert \gg 1$, which can
be achieved through a large input power $\wp _{j}$ \cite{Vitali(1)}. So, the
second-order product fluctuations terms $\delta c_{j}^{\dag }\delta c_{j}$, $%
\delta c_{j}\delta b_{j}$ and $\delta c_{j}\delta b_{j}^{\dag }$ can be
safely ignored. Hence, we get
\begin{eqnarray}
\delta \dot{b}_{j} &=&-\left( \gamma _{j}/2+i\omega _{\mu _{j}}\right)
\delta b_{j}+\mathcal{G}_{j}\left( \delta c_{j}-\delta c_{j}^{\dag }\right) +%
\sqrt{\gamma _{j}}\zeta _{j}^{in},  \label{E11} \\
\delta \dot{c}_{j} &=&-\left( \kappa _{j}/2-i\Delta _{j}^{\prime }\right)
\delta c_{j}-\mathcal{G}_{j}\left( \delta b_{j}^{\dag }+\delta b_{j}\right) +%
\sqrt{\kappa _{j}}\delta \digamma _{j}^{in},  \label{E12}
\end{eqnarray}%
$\mathcal{G}_{j}$ $=g_{j}\left\vert c_{js}\right\vert =g_{j}\sqrt{\bar{n}%
_{cav}^{j}}$ and $\bar{n}_{cav}^{j}$ are respectively the $j^{th}$
many-photon optomechanical coupling and the average number of photons
circulating inside the $j^{th}$ cavity \cite{Marquardt}. We note that Eqs. (%
\ref{E11})-(\ref{E12}) are obtained by choosing the phase $\varphi_{j}$ of
the $j^{th}$ input laser field as $\varphi _{j}=-\arctan(2\Delta
_{j}^{\prime }/\kappa _{j})$ or equivalently $c_{js}=-i\left\vert
c_{js}\right\vert $. Moving into an interaction picture by introducing the
operators $\delta \tilde{b}_{j}=\delta b_{j}e^{i\omega _{\mu _{j}}t}$ and $%
\delta \tilde{c}_{j}=\delta c_{j}e^{-i\Delta _{j}^{\prime }t}$, Eqs. (\ref%
{E11})-(\ref{E12}) become
\begin{eqnarray}
\delta \dot{\tilde{b}}_{j} &=&-\frac{\gamma _{j}}{2}\delta \tilde{b}_{j}\ +%
\mathcal{G}_{j}\left( \delta \tilde{c}_{j}e^{i\left( \Delta _{j}^{\prime
}+\omega _{\mu _{j}}\right) t}-\delta \tilde{c}_{j}^{\dag }e^{-i\left(
\Delta _{j}^{\prime }-\omega _{\mu _{j}}\right) t}\right) +\sqrt{\gamma _{j}}%
\tilde{\zeta}_{j}^{in},  \label{E13} \\
\delta \dot{\tilde{c}}_{j} &=&-\frac{\kappa _{j}}{2}\delta \tilde{c}_{j}-%
\mathcal{G}_{j}\left( \delta \tilde{b}_{j}e^{-i\left( \Delta _{j}^{\prime
}+\omega _{\mu _{j}}\right) t}+\delta \tilde{b}_{j}^{\dag }e^{-i\left(
\Delta _{j}^{\prime }-\omega _{\mu _{j}}\right) t}\right) +\sqrt{\kappa _{j}}%
\delta \tilde{\digamma}_{j}^{in}.  \label{E14}
\end{eqnarray}%
Next, we focus on the red sideband,i.e., $\Delta _{j}^{\prime }=-\omega
_{\mu _{j}}$, which is relevant for quantum states transfer \cite{steering}.
In addition, in the resolved-sideband regime, where $\omega _{\mu _{j}}\gg
\kappa _{j}$, one can perform the rotating wave approximation (RWA) \cite%
{SGHofer}, which allows us to drop fast-rotating terms oscillating with $\pm
2\omega _{\mu _{j}}$ in Eqs. (\ref{E13})-(\ref{E14}). We then obtain
\begin{eqnarray}
\quad \delta \dot{\tilde{b}}_{j} &=&-\frac{\gamma _{j}}{2}\delta \tilde{b}%
_{j}+\mathcal{G}_{j}\delta \tilde{c}_{j}+\sqrt{\gamma _{j}}\tilde{\zeta}%
_{j}^{in},  \label{E15} \\
\delta \dot{\tilde{c}}_{j} &=&-\frac{\kappa _{j}}{2}\delta \tilde{c}_{j}-%
\mathcal{G}_{j}\delta \tilde{b}_{j}+\sqrt{\kappa _{j}}\delta \tilde{\digamma}%
_{j}^{in}.  \label{E16}
\end{eqnarray}

\subsection{Covariance matrix}

Equations (\ref{E15})-(\ref{E16}) can be solved in the frequency domain
using the Fourier transform of each equation. So, one obtains
\begin{eqnarray}
\delta \tilde{c}_{j}(\omega ) &=&-\frac{\mathcal{G}_{j}}{\Xi _{j}(\omega )}%
\sqrt{\gamma _{j}}~\tilde{\zeta}_{j}^{in}(\omega )+\frac{\left( \gamma
_{j}/2+i\omega \right) }{\Xi _{j}(\omega )}\sqrt{\kappa _{j}}~\delta \tilde{%
\digamma}_{j}^{in}(\omega ),  \label{deltactilde} \\
\delta \tilde{b}_{j}(\omega ) &=&\frac{\left( \kappa _{j}/2+i\omega \right)
}{\Xi _{j}(\omega )}\sqrt{\gamma _{j}}~\tilde{\zeta}_{j}^{in}(\omega )+\frac{%
\mathcal{G}_{j}}{\Xi _{j}(\omega )}\sqrt{\kappa _{j}}~\delta \tilde{\digamma}%
_{j}^{in}(\omega ),  \label{deltabtilde}
\end{eqnarray}%
with $\Xi _{j}(\omega )=\mathcal{G}_{j}^{2}+\left( \gamma _{j}/2+i\omega
\right) \left( \kappa _{j}/2+i\omega \right) $. \newline
The correlation function of any pair of fluctuation operators is acquired as
\begin{equation}
\mathcal{V}_{kk^{\prime }}=\frac{1}{4\pi ^{2}}\int \int d\omega d\omega
^{\prime }e^{-i(\omega +\omega ^{\prime })t}\mathcal{V}_{kk^{\prime
}}(\omega ,\omega ^{\prime }),  \label{CM elements}
\end{equation}%
$\mathcal{V}_{kk^{\prime }}(\omega ,\omega ^{\prime })=\langle \tilde{u}%
_{k}(\omega )\tilde{u}_{k^{\prime }}(\omega ^{\prime })+\tilde{u}_{k^{\prime
}}(\omega ^{\prime })\tilde{u}_{k}(\omega )\rangle /2$ (for $k,k^{\prime
}=1,..,8$) is the frequency-domain correlation function between elements $k$
and $k^{\prime }$ of $\tilde{u}=(\delta \tilde{q}_{1}^{m},\delta \tilde{p}%
_{1}^{m},\delta \tilde{q}_{2}^{m},\delta \tilde{p}_{2}^{m},\delta \tilde{q}%
_{1}^{op},\delta \tilde{p}_{1}^{op},\delta \tilde{q}_{2}^{op},\delta \tilde{p%
}_{2}^{op})$, where $\delta \tilde{q}_{j}^{m}=(\delta \tilde{b}_{j}^{\dagger
}+\delta \tilde{b}_{j})/\sqrt{2}$ and $\delta \tilde{p}_{j}^{m}=i(\delta
\tilde{b}_{j}^{\dagger }-\delta \tilde{b}_{j})/\sqrt{2}$ \Big($\delta \tilde{%
q}_{j}^{op}=(\delta \tilde{c}_{j}^{\dag }+\delta \tilde{c}_{j})/\sqrt{2}$
and $\delta \tilde{p}_{j}^{op}=i(\delta \tilde{c}_{j}^{\dag }-\delta \tilde{c%
}_{j})/\sqrt{2}$\Big) are the quadratures position and momentum of the $%
j^{th}$ mechanical(optical) mode. Using Eq. (\ref{CM elements}) and the
correlation properties of the noise operators given by Eqs. [(\ref{E4-})-(%
\ref{E9})], the covariance matrix (CM) of the whole system (four-mode
Gaussian states) can be evaluated explicitly. In this work, we are
interested in the non-classical properties of the steady state of two-mode
Gaussian states. So, we shall focus onto the entanglement and the quantum
discord of two different bipartite subsystems, which can be formed by
tracing over the remaining degree of freedom. \newline
The $4\times 4$ CM $\mathcal{V}^{i}$ of the selected bipartition that
obtained by neglecting the rows and columns of the uninteresting mode, can
be written in the following standard form with diagonal sub-blocks,
\begin{equation}
\mathcal{V}^{i}=\left(
\begin{array}{cc}
\mathcal{V}_{1}^{i} & \mathcal{V}_{3}^{i} \\
\mathcal{V}_{3}^{i\mathrm{T}} & \mathcal{V}_{2}^{i}%
\end{array}%
\right) ,\text{ for }i \in \{m,op\}\text{,}  \label{CM}
\end{equation}%
with $\mathcal{V}_{1}^{i}=\mathrm{diag}(\nu _{1}^{i},\nu _{1}^{i})$, $%
\mathcal{V}_{2}^{i}=\mathrm{diag}(\nu _{2}^{i},\nu _{2}^{i})$ and $\mathcal{V%
}_{3}^{i}=\mathrm{diag}(\nu _{3}^{i},\nu _{3}^{^{\prime }i})$, where $%
i\equiv m$ ($op$) refers the two-mode mechanical(optical) subsystem. The
matrix $\mathcal{V}^{i}$ is a real, symmetric and positive definite matrix.
The $2\times 2$ matrices $\mathcal{V}_{1}^{m}$ and $\mathcal{V}_{2}^{m}$($%
\mathcal{V}_{1}^{op}$ and $\mathcal{V}_{2}^{op}$) represent the first and
second mechanical(optical) modes, while the correlations between them are
described by $\mathcal{V}_{3}^{m}$($\mathcal{V}_{3}^{op}$). Being a physical
state, the $\mathcal{V}^{i}$-(CM) should satisfy the Heisenberg-Robertson
uncertainty principle \ $\mathcal{V}^{i}+i\Omega _{3}/2\geqslant 0$, where $%
\Omega _{3}=\oplus _{1}^{3}i\sigma _{y}$ and $\sigma _{y}$ are respectively
the so-called symplectic matrix and the $y$-Pauli matrix \cite{SSM}. \newline
Considering identical cavities, i.e., $\omega _{\mu _{1,2}}=\omega _{\mu }, $
$\gamma _{1,2}=\gamma ,$ $\kappa _{1,2}=\kappa $, etc.., the elements of the
matrices $\mathcal{V}^{i}$ ($i\in \{m,op\}$) defined by Eq.(\ref{CM}) are
\begin{eqnarray}
\nu _{1}^{m} &=&\nu _{2}^{m}=\frac{\left( 2n_{th}+1\right) \left( 1+\Gamma
+\Gamma \mathcal{C}\right) +\mathcal{C}\cosh 2r}{2(1+\Gamma )\left( 1+%
\mathcal{C}\right) },\text{ \ }\nu _{3}^{m}=-\nu _{3}^{^{\prime }m}=\frac{%
\mathcal{C}\sinh 2r}{2(1+\Gamma )\left( 1+\mathcal{C}\right) },\text{\ \ \ \
\ \ \ \ }  \label{a} \\
\text{ \ \ \ \ \ \ \ \ \ \ \ \ \ }\nu _{1}^{op} &=&\nu _{2}^{op}=\frac{%
\left( 2n_{th}+1\right) \Gamma \mathcal{C}+\left( 1+\Gamma +\mathcal{C}%
\right) \cosh 2r}{2(1+\Gamma )\left( 1+\mathcal{C}\right) },\text{ }\nu
_{3}^{op}=-\nu _{3}^{^{\prime }op}=\frac{\left( 1+\Gamma +\mathcal{C}\right)
\sinh 2r}{2(1+\Gamma )\left( 1+\mathcal{C}\right) },  \label{c}
\end{eqnarray}%
where $\Gamma $ $=\gamma /\kappa $ and $\mathcal{C}=4\mathcal{G}^{2}/\gamma
\kappa =\frac{8\omega _{c}^{2}\wp }{\gamma m_{\mu }\omega _{\mu }\omega
_{L}l^{2}\left[ \left( \frac{\kappa }{2}\right) ^{2}+\omega _{_{\mu }}^{2}%
\right] }$ are respectively the damping ratio and the optomechanical
cooperativity \cite{OET}. \newline
In the following, using the approach sketched above, one can analysis both
entanglement and quantum discord in the mechanical and optical subsystems.
For this, we shall use the Gaussian R\'{e}nyi-$2$ entropy \cite{AGS}.

\section{Gaussian R\'{e}nyi-2 quantum correlations \label{sec3}}

R\'{e}nyi-$\alpha $ entropies were first introduced by Alfred R\'{e}nyi as a
generalisation of the usual concept of entropy \cite{Renyi}. These entropies
encompass several entropic measures, including the Shannon entropy and the
von Neumann entropy \cite{AGS}. Motivated by the mathematical foundations of
entropy, the R\'{e}nyi-$\alpha $ entropies of a given quantum state $%
\mathcal{\varrho }$ are defined as \cite{AGS}
\begin{equation}
\mathcal{S}_{\alpha }\mathcal{(\varrho )\equiv (}1-\alpha \mathcal{)}%
^{-1}\ln \mathrm{Tr}\left( \varrho ^{\alpha }\right) ,  \label{Renyi-alpha}
\end{equation}%
where $\alpha \in (0,1)\cup (1,+\infty )$, which usually written as $\alpha
\in (0,1,+\infty )$ \cite{Berta2015}. These entropies are continuous,
non-negatives, invariants under the action of the unitary operations,
additive on tensor-product states, and reduce in the limit $\alpha
\rightarrow 1$ to the conventional von Neumann entropy defined by $\mathcal{%
S(\varrho )=}-\mathrm{Tr}\left( \varrho \ln \varrho \right) $ \cite{Martin}.
In quantum information theory, this famous entropy captures the degree of
quantum information possessed in an ensemble of a large number of
independent and identically distributed (i.i.d.) copies of the state \cite%
{Schumacher}, where it has been intensively used in different fields of
sciences such as communication and coding theory \cite{CCT}, statistical
physics \cite{Sta}, quantum computation \cite{VVedral}, statistics and
related fields \cite{Statistique}. In addition, the von Neumann entropy has
been widely used with the well known name \textit{entropy of entanglement}
to quantify entanglement in miscellaneous areas of physics like theory of
black holes \cite{DKT}, relativistic quantum field theory \cite{QFT}, spins
of noninteracting electron gases \cite{Oh}, three-level atom under influence
of a Kerr-like medium \cite{Aty}, harmonic lattice systems \cite{P and E},
the Lipkin-Meshkov-Glick model \cite{Rico}, and also spin-orbital models
\cite{Horsch}. We notice that in the case of \textit{pure states}, many
entanglement measures (including entanglement cost \cite{Wootters,VDC},
entanglement of formation (EoF) \cite{Wootters,EoF}, squashed entanglement
\cite{Christandl}, and also the quantum discord \cite{G(1),ZA}) coincide
with the entropy of entanglement as the only measure of entanglement in this
case \cite{HorodeckiS}. In contrast, some other quantifiers of entanglement
such as the negativity \cite{Zyc} and the logarithmic negativity \cite%
{MBPlenio} can not coincide with the entropy of entanglement even on pure
states. \newline
On the other hand, for $\alpha \neq 1$, the R\'{e}nyi-$\alpha $ entropies
are relevant in situations other than the i.d.d. setting \cite{Tomamichel}.
For example, they are useful to characterize many tasks of information
processing with either a single or finite number of resource utilizations
\cite{Hayashi}, and also to establish strong converse theorems \cite{Konig}.
Now, replacing $\alpha $ by the special value $2$ in Eq. (\ref{Renyi-alpha}%
), we obtain the simple expression $\mathcal{S}_{2}\mathcal{(\varrho )=}-\ln
\mathrm{Tr}\left( \varrho ^{2}\right) $ corresponding to the R\'{e}nyi-2
entropy \cite{AGS}. It has been proven that for arbitrary Gaussian states,
the R\'{e}nyi-2 entropy fulfills the strong subadditivity inequality; $%
\mathcal{S}_{2}\left( \mathcal{\varrho }_{AB}\right) +\mathcal{S}_{2}\left(
\mathcal{\varrho }_{BC}\right) \geqslant \mathcal{S}_{2}\left( \mathcal{%
\varrho }_{ABC}\right) +\mathcal{S}_{2}\left( \mathcal{\varrho }_{B}\right) $
\cite{AGS}, which made it possible to define bona fide Gaussian measures of
entanglement \cite{G(1),AL} and quantum discord \cite{Kavan Modi,G(1),ZA}.

\subsection{ Gaussian R\'{e}nyi-2 entanglement}

For generally mixed two-mode Gaussian states $\mathcal{\varrho }%
_{A^{i}B^{i}} $, the Gaussian R\'{e}nyi-2 entanglement (GR2E) $\mathcal{E}%
_{2}$ admits a complicated parametric formula \cite{AGS, AL}, that can be
computed numerically only by programming \cite{AGS, AL}. However, for all
two-mode Gaussian pure states as well as for several important classes of
mixed states (including squeezed thermal states STS \cite{G(1)}), closed
formulas for the GR2E have been obtained \cite{AGS,AL}. \newline
On one hand, the covariance matrix $\mathcal{V}^{i}$ (\ref{CM}) is in the
standard form \cite{G(1)}. On the other hand, the $\mathcal{V}_{3}^{i}$%
-block matrix has the form $\mathcal{V}_{3}^{i}=\mathrm{diag}(\nu
_{3}^{i},-\nu _{3}^{i})$ (see Eqs. [(\ref{CM})-(\ref{c})]), which
corresponds to the squeezed-thermal states STS \cite{G(1)}. Thus, the GR2E $%
\mathcal{E}_{2}^{i}$ ($i\in \{m,op\}$) for two-mode Gaussian states with CM $%
\mathcal{V}^{i}$ (\ref{CM}) is given by \cite{AL}
\begin{equation}
\mathcal{E}_{2}^{i}=\left\{
\begin{array}{c}
\frac{1}{2}\ln \left[ \frac{(4f^{i}+1)s^{i}-\sqrt{\left[ (4f^{i}-1)^{2}-16%
\left( d^{i}\right) ^{2}\right] \left[ \left( s^{i}\right) ^{2}-\left(
d^{i}\right) ^{2}-f^{i}\right] }}{4(\left( d^{i}\right) ^{2}+f^{i})}\right]
^{2}\text{if \ \ \ \ }4|d^{i}|+1\leq 4f^{i}<4s^{i}-1, \\
0\text{ \ \ \ \ \ \ \ \ \ \ \ \ \ \ \ \ \ \ \ \ \ \ \ \ \ \ \ \ \ \ \ \ \ \
\ \ \ \ \ \ \ \ \ \ \ \ \ \ \ \ \ \ \ \ \ \ \ \ \ \ \ \ \ if \ \ \ \ \ \ }
4f^{i}\geqslant 4s^{i}-1,%
\end{array}%
\right.  \label{-E33-}
\end{equation}%
where $s^{i}=\left( \nu _{1}^{i}+\nu _{2}^{i}\right) /2$, $d^{i}=\left( \nu
_{1}^{i}-\nu _{2}^{i}\right) /2$ and $f^{i}=\sqrt{\det \mathcal{V}^{i}}$ for
$i\in \{m,op\}$.\newline
We note that, another computable quantifier of entanglement in two-mode
Gaussian states called Gaussian intrinsic entanglement (GIE), was introduced
more recently by Mi\u{s}ta and Tatham \cite{Tatham}. It has been \textit{%
surprisingly} found that GIE coincides exactly with the GR2E \cite{Tatham}
for all pure two-mode Gaussian states as well as for relevant subclasses of
Gaussian states, including symmetric states, squeezed thermal states, and
so-called GLEMS Gaussian states of partial minimum uncertainty \cite{AL}%
,leading to conjecture that GIE and GR2E are equal on all Gaussian states
\cite{Tatham}. \newline
\begin{figure}[t]
\centerline{\includegraphics[width=0.5\columnwidth,height=4.5cm]{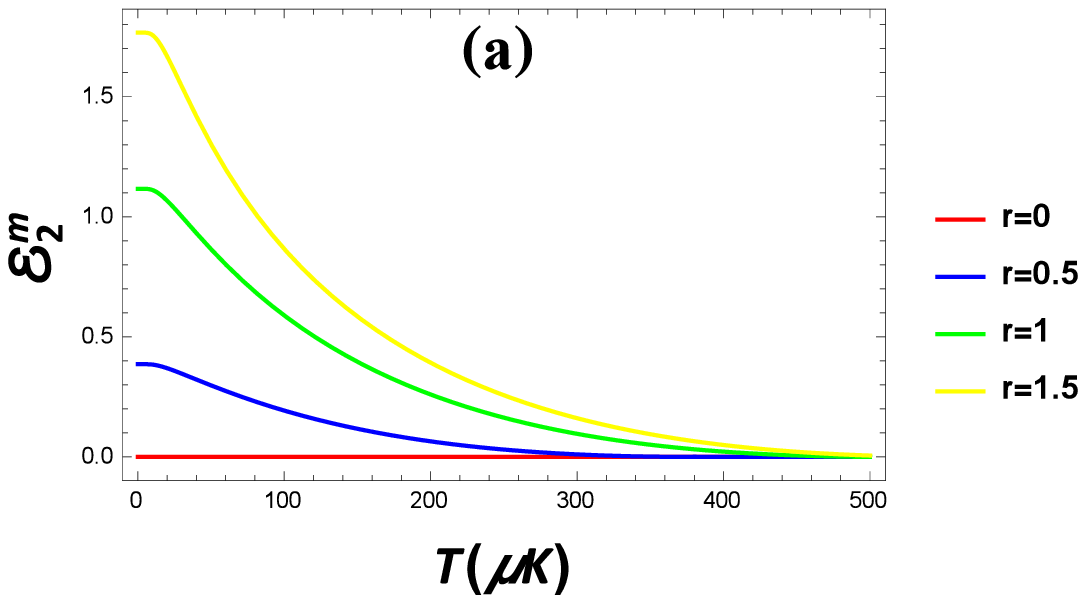}%
\includegraphics[width=0.5\columnwidth,height=4.5cm]{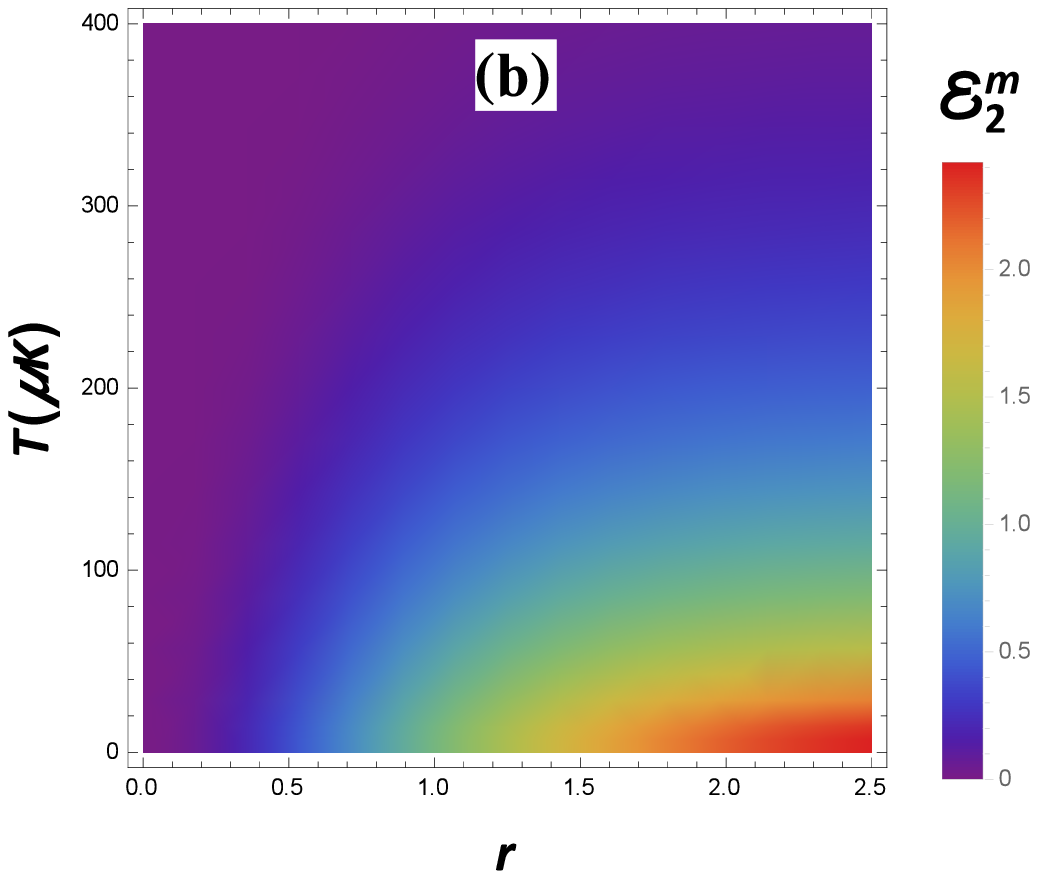}} %
\centerline{\includegraphics[width=0.5\columnwidth,height=4.5cm]{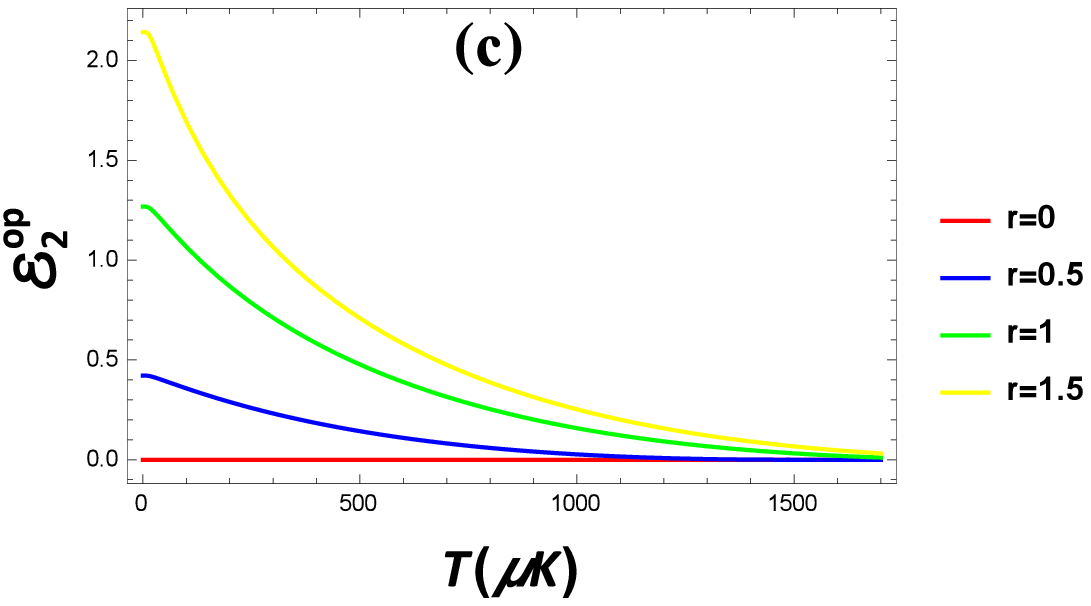}%
\includegraphics[width=0.5\columnwidth,height=4.5cm]{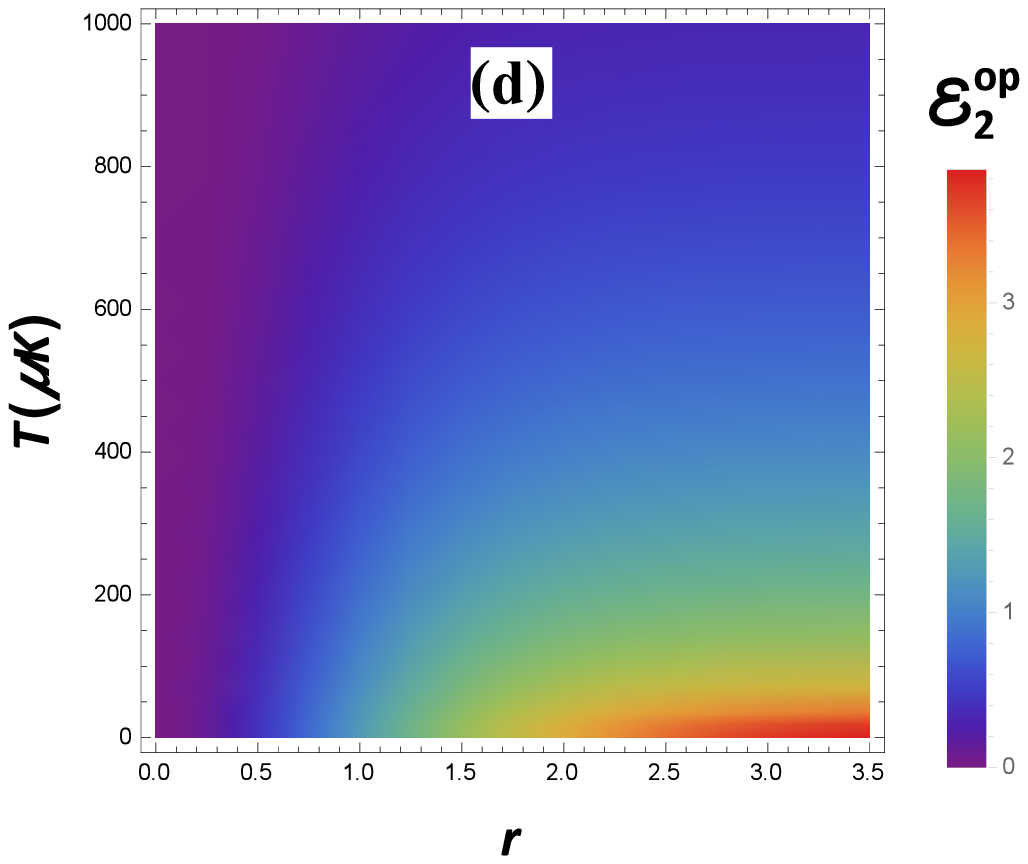}}
\caption{GR2E $\mathcal{E}_{2}^{i}$ versus the thermal bath temperature $T$
for various squeezing $r$. [(a)-(b)] GR2E $\mathcal{E}_{2}^{m}$ of the two
mechanical modes, [(c)-(d)] GR2E $\mathcal{E}_{2}^{op}$ of the two optical
modes. The optomechanical cooperativity $\mathcal{C}$ and the damping ratio $%
\Gamma =\frac{\protect\gamma }{\protect\kappa }$ are respectively $\mathcal{C%
}=34$ and $\Gamma =\frac{\protect\gamma }{\protect\kappa }\approx 0.01$. }
\label{fig.2}
\end{figure}
In our simulations, we have used parameters taken from experiment \cite%
{Groblacher}. As mentioned above, we take identical mechanical modes with
frequency $\omega _{\mu }=2\pi \times 947~\mathrm{KHz}$, mass $m_{\mu} =145~%
\mathrm{ng}$, and damping rate $\gamma =2\pi \times 140~\mathrm{Hz}$, and
identical cavities having length $l=25~\mathrm{\ mm}$, frequency $\omega
_{c}=2\pi \times 5.26\times 10^{14}~\mathrm{Hz}$, decay rate $\kappa =2\pi
\times 172$ $\mathrm{KHz}$ and pumped by laser fields of frequency $\omega
_{L}=2\pi \times 2.82\times 10^{14}$ $\mathrm{Hz}$ and power $\wp =1.5~%
\mathrm{mW}$ (for the parameters $\kappa $ and $\wp $, see the supplementary
information in \cite{Groblacher}). We note that, the situation $\omega _{\mu
}\gg \kappa $ ($\omega _{\mu }=2\pi \times 947~\mathrm{KHz}$, $\kappa =2\pi
\times 172~\mathrm{KHz}$) corresponding to the resolved sideband regime \cite%
{SGHofer}, justifies the use of the RWA in section \ref{sec2}. Finally, the
values of the squeezing parameter $r$ are chosen of the same order of
magnitude as in \cite{JieLi}. \newline
Fig. \ref{fig.2} shows that for a fixed value of the thermal bath
temperature $T$, the GR2E $\mathcal{E}_{2}^{m}$($\mathcal{E}_{2}^{op}$) of
the mechanical(optical) modes increases as the squeezing parameter $r$
increases, which means that the squeezing enhances the quantum correlations
in the two considered subsystems. On the other hand, in the case of no
injection of the squeezed light ($r=0$), we remark that $\mathcal{E}%
_{2}^{op}=\mathcal{E}_{2}^{m}=0$, which indicates that the two
non-interacting mechanical(optical) modes cannot be entangled without
squeezed light. However, for $r\neq 0$, one can see that both mechanical and
optical entanglement are detectable, meaning that there are \textit{%
inseparable quantum correlations} between the two mechanical(optical) modes,
even though they are spatially disjoint. These results can be explained by
two types of quantum fluctuations transfer. The first one from the two-mode
squeezed light to the optical subsystem, which corresponds to \textit{%
light-light quantum fluctuations transfer}, and the second one from the
two-mode squeezed light to the mechanical subsystem, which corresponds to
\textit{light-matter quantum fluctuations transfer} \cite{discord}. Fig. \ref%
{fig.2} shows also, that the entanglement $\mathcal{E}_{2}^{m}$ and $%
\mathcal{E}_{2}^{op}$ decay rapidly to zero when the thermal bath
temperature $T$ increases, remarking that the mechanical entanglement $%
\mathcal{E}_{2}^{m}$ is more affected by thermal effects than the optical
entanglement $\mathcal{E}_{2}^{op}$. In fact, for a given squeezing $r\neq 0$%
, Fig. \ref{fig.2} shows that always $T_{c}^{op}>T_{c}^{m},$ where $%
T_{c}^{op}$($T_{c}^{m}$) is the optical(mechanical) critical temperature
defined by $T\geqslant T_{c}^{i}\Rightarrow \mathcal{E}_{2}^{i}=0$ ($i\in
\{m,op\}$), from which the entangled optical(mechanical) modes become
separable. This can be well understood relying on the recent work \cite%
{YaoYao}. Indeed, the two optical modes are subject to the zero-point
fluctuation from their vacuum environments, while the mechanical modes are
subject to the thermal noise from their baths. Consequently, the two optical
modes, which are in a more coherent environment than that of the mechanical
modes, should have a robust entanglement with larger degree, knowing that
coherences are a genuine resource of quantum entanglement \cite{C and E}.%
\newline
\begin{figure}[t]
\centerline{\includegraphics[width=0.5\columnwidth,height=4.5cm]{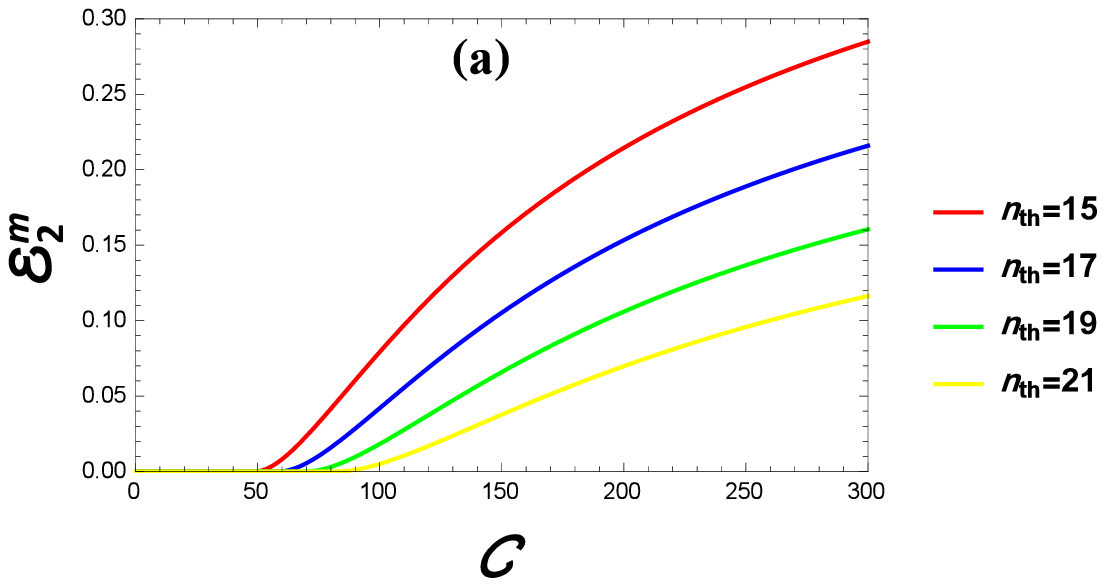}%
\includegraphics[width=0.5\columnwidth,height=4.5cm]{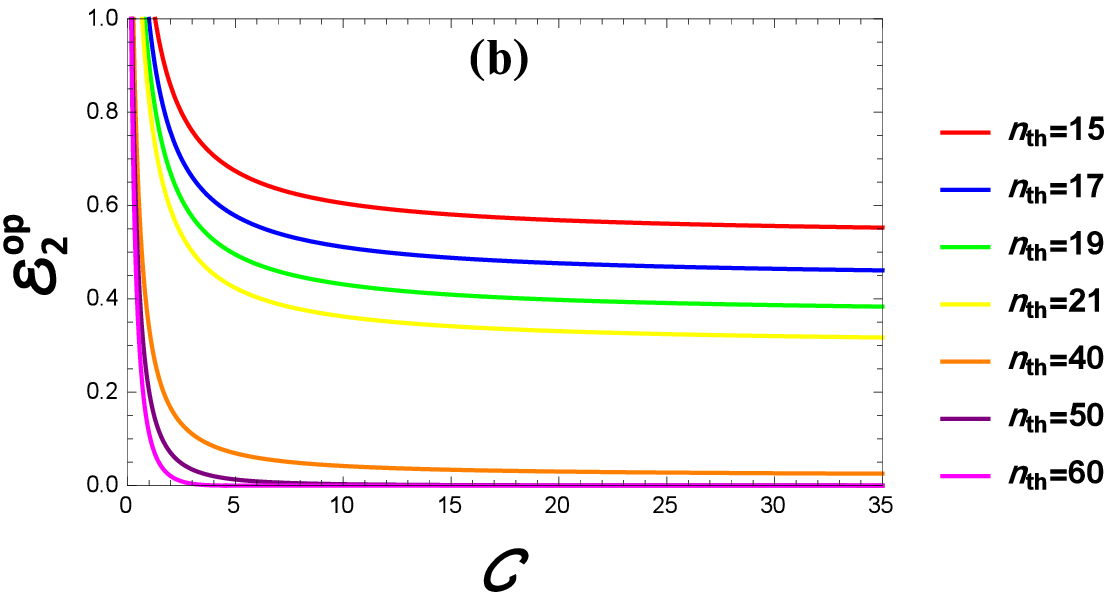}}
\caption{GR2E $\mathcal{E}_{2}^{i}$ versus the dimensionless optomechanical
cooperativity $\mathcal{C}$ for various values of the mean thermal photons
number $n_{th}$. (a) Mechanical GR2E $\mathcal{E}_{2}^{m}$, (b) optical GR2E
$\mathcal{E}_{2}^{op}$. The squeezing degree $r$ and the damping ratio $%
\Gamma =\frac{\protect\gamma }{\protect\kappa }$ are respectively $r=2$ and $%
\Gamma =\frac{\protect\gamma }{\protect\kappa }\approx 0.01$.}
\label{fig.3}
\end{figure}
Next, fixing the squeezing parameter as $r=2$, Fig. \ref{fig.3} shows the
behavior of the GR2E $\mathcal{E}_{2}^{m}$ and $\mathcal{E}_{2}^{op}$ versus
the optomechanical cooperativity $\mathcal{C}$ for different values of the
mean thermal photons number $n_{\mathrm{th}}$. Unlike the optical
entanglement $\mathcal{E}_{2}^{op}$ illustrated in Fig. \ref{fig.3}(b), Fig. %
\ref{fig.3}(a) shows that non-zero optomechanical coupling ($\mathcal{C}\neq
0$) is required to entangle the two mechanical modes. This can be understood
using Eq. (\ref{a}), where $\det \mathcal{V}_{3}^{m}=0$ when $\mathcal{C}=0$%
, knowing that $\det \mathcal{V}_{3}^{i}<0$ ($i\in \{m,op\}$) is a necessary
condition for a two-mode Gaussian state to be entangled\ \cite{Keller2}. In
addition, Fig. \ref{fig.3} shows that the entanglement $\mathcal{E}_{2}^{m}$
and $\mathcal{E}_{2}^{op}$ behave oppositely under the same circumstances.
Indeed, for a given value of the mean thermal photons number $n_{\mathrm{th}}
$, it is clear that when the optomechanical cooperativity $\mathcal{C}$
increases, the mechanical entanglement $\mathcal{E}_{2}^{m}$ increases,
while the optical entanglement $\mathcal{E}_{2}^{op}$ decreases. This can be
explained by the quantum fluctuations transfer from the optical subsystem to
the mechanical subsystem (\textit{light-matter quantum fluctuations transfer}%
). Finally, one can see also from Fig. \ref{fig.3} that the optical
subsystem requires high values of $n_{\mathrm{th}}$ to switch from entangled
to separable states (see Fig. \ref{fig.3}(b)), which reflects the robust
behavior of the optical modes versus thermal noises. This is consistent with
the results obtained in Fig. \ref{fig.2}.

\subsection{Gaussian R\'{e}nyi-2 discord}

We now turn our attention to the quantification of quantum correlations
beyond entanglement which are commonly known as \textit{the quantumness of
correlations} \cite{ZA}. As mentioned previously, we will use the Gaussian R%
\'{e}nyi-2 discord (GR2D) $\mathcal{D}_{2}^{i}(\varrho _{A^{i}/B^{i}})$
defined in \cite{AGS}, as the difference between mutual information $%
\mathcal{I}_{2}(\varrho _{A^{i}/B^{i}})$ and classical correlations $%
\mathcal{J}_{2}(\varrho _{A^{i}/B^{i}})$
\begin{equation}
\mathcal{D}_{2}^{i}(\varrho _{A^{i}/B^{i}})\doteq \mathcal{I}_{2}(\varrho
_{A^{i}:B^{i}})-\mathcal{J}_{2}(\varrho _{A^{i}/B^{i}})\text{ for }i\in
\{m,op\}\text{.}  \label{GRQD}
\end{equation}%
The GR-2 mutual information $\mathcal{I}_{2}(\varrho _{A^{i}:B^{i}})$ for an
arbitrary bipartite Gaussian state $\varrho _{A^{i}B^{i}}$ is given by $%
\mathcal{I}_{2}(\varrho _{A^{i}:B^{i}})=\mathcal{S}_{2}\mathcal{(}\varrho
_{A^{i}}\mathcal{)+S}_{2}\mathcal{(}\varrho _{B^{i}}\mathcal{)-S}_{2}%
\mathcal{(}\varrho _{A^{i}B^{i}}\mathcal{)}$, where $\varrho _{A^{i}}$ and $%
\varrho _{B^{i}}$ are the two marginals of $\varrho _{A^{i}B^{i}}$, while a
GR-2 measure of one-way classical correlations $\mathcal{J}_{2}(\varrho
_{A^{i}/B^{i}})$ can be obtained as the maximum decrease in the R\'{e}nyi-2
entropy of subsystem $A^{i}$, given that a Gaussian measurement has been
performed on the subsystem $B^{i}$, where the maximization is over all
Gaussian measurements, which map Gaussian states into Gaussian states \cite%
{G(1)}.\newline
Interestingly, we note that Gaussian states have been shown to play a
privileged role in quantum optics and bosonic field theories, essentially
since the very early steps of such theories, partly due to the associated
reduced mathematical complexity, and partly due to their immense importance
for quantum information processing and their easy experimental production
and detection \cite{Lami,Madsen}.\newline
\begin{figure}[t]
\centerline{\includegraphics[width=0.5\columnwidth,height=4.5cm]{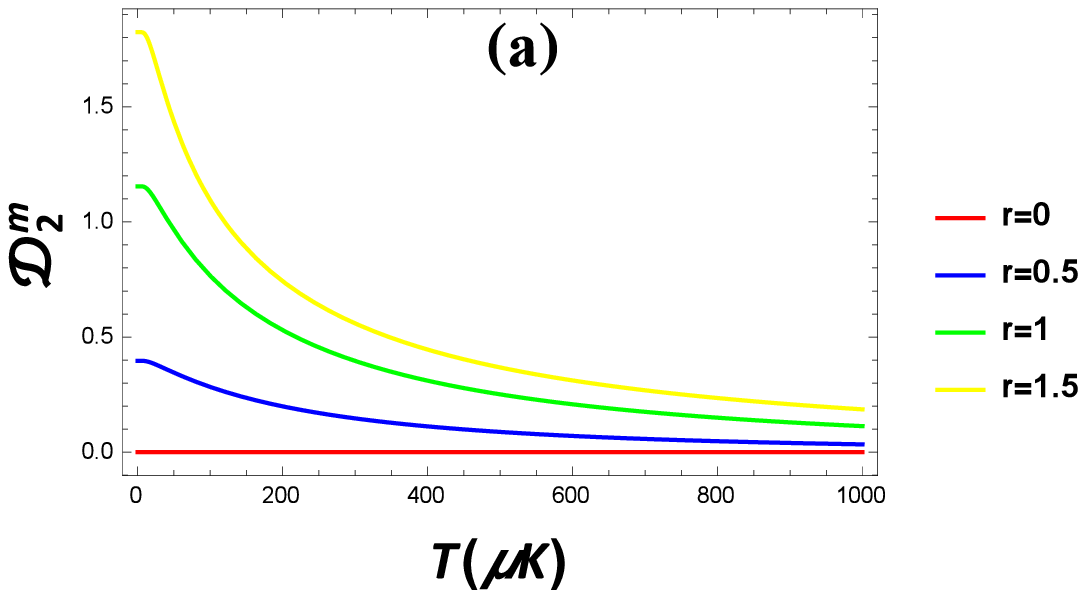}%
\includegraphics[width=0.5\columnwidth,height=4.5cm]{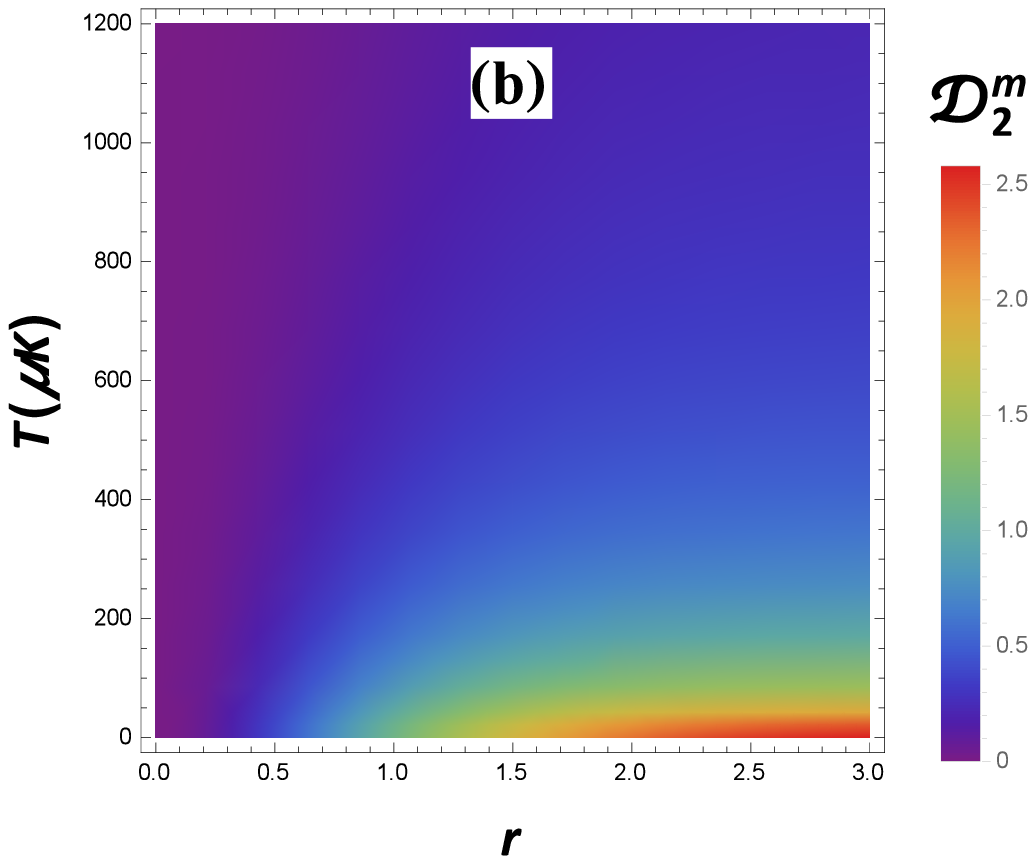}} %
\centerline{\includegraphics[width=0.5\columnwidth,height=4.5cm]{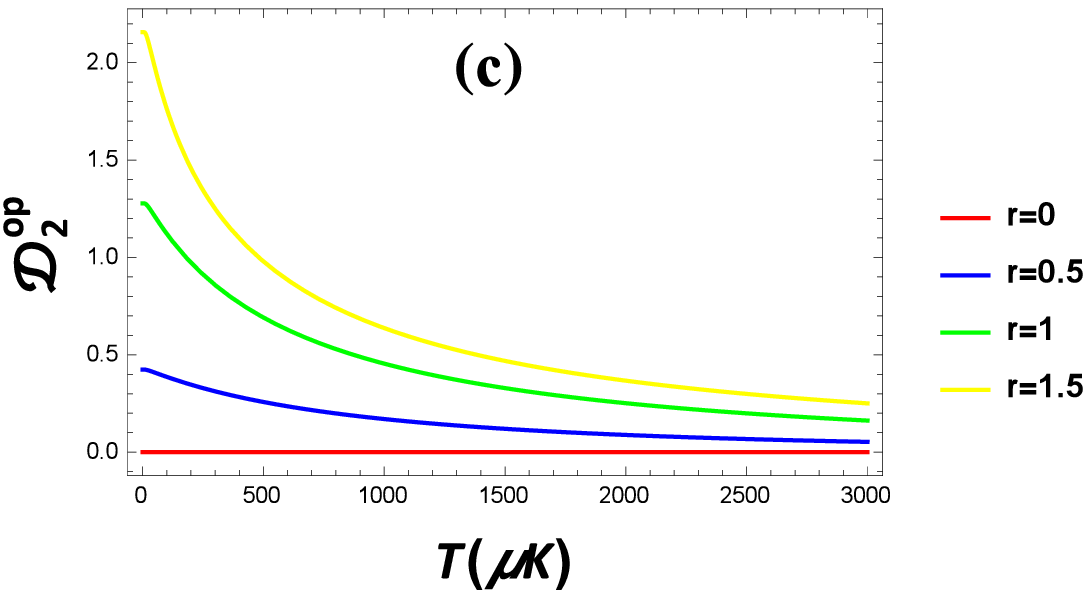}%
\includegraphics[width=0.5\columnwidth,height=4.5cm]{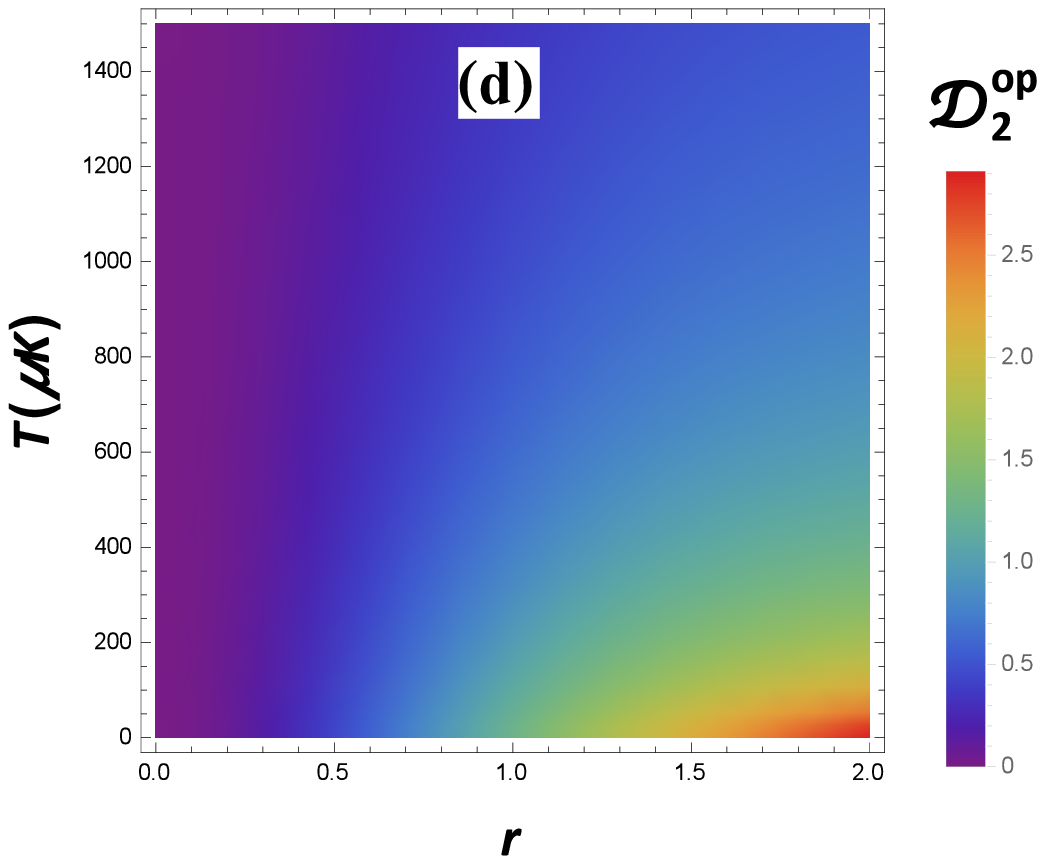}}
\caption{GR2D $\mathcal{D}_{2}^{i}$ versus the thermal bath temperature $T$
for various values of the squeezing parameter $r$. [(a)-(b)] Mechanical GR2D
$\mathcal{D}_{2}^{m}$, [(c)-(d)] optical GR2D $\mathcal{D}_{2}^{op}$. The
optomechanical cooperativity $\mathcal{C}$ and the damping ratio $\Gamma =%
\frac{\protect\gamma }{\protect\kappa }$ are the same as in Fig. \protect\ref%
{fig.2}.}
\label{fig.4}
\end{figure}
For a two-mode Gaussian state $A^{i}$ and $B^{i}$ with CM (\ref{CM}), the R%
\'{e}nyi-2 measure of one-way quantum discord (\ref{GRQD}) admits the
following expression \cite{AGS}
\begin{equation}
\mathcal{D}_{2}^{i}(\varrho _{A^{i}/B^{i}})=\ln (2\nu _{2}^{i})-\frac{1}{2}%
\ln \left( 16\det \mathcal{V}^{i}\right) +\frac{1}{2}\ln \left( \epsilon
^{i}\right) \text{ for }i\in \{m,op\}\text{,}  \label{GR2QD}
\end{equation}%
with
\begin{equation}
\epsilon ^{i}=\left\{
\begin{array}{c}
4\nu _{1}^{i}\left( \nu _{1}^{i}\text{ }-\frac{\left( \nu _{3}^{i}\right)
^{2}}{\nu _{2}^{i}}\right) \text{\ \ if\ \ }\left[ 4\nu _{1}^{i}\left( \nu
_{2}^{i}\right) ^{2}\left( \nu _{3}^{^{\prime }i}\right) ^{2}-\left( \nu
_{3}^{i}\right) ^{2}\left( \nu _{1}^{i}+4\nu _{2}^{i}\left( \nu
_{3}^{^{\prime }i}\right) ^{2}\right) \right] \text{ }\times \\
\left[ 4\nu _{1}^{i}\ (\nu _{2}^{i})^{2}\left( \nu _{3}^{i}\right) ^{2}-\ \
\left( \nu _{3}^{^{\prime }i}\right) ^{2}\left( \nu _{1}^{i}+4\nu
_{2}^{i}\left( \nu _{3}^{i}\right) ^{2}\right) \right] <0, \\
\left[ \frac{4\left\vert \nu _{3}^{i}\nu _{3}^{^{\prime }i}\right\vert +2%
\sqrt{\left[ \nu _{1}^{i}\left( 4(\nu _{2}^{i})^{2}-1\right) -4\nu
_{2}^{i}(\nu _{3}^{^{\prime }i})^{2}\right] \left[ \nu _{1}^{i}\left[ 4(\nu
_{2}^{i})^{2}-1\right] -4\nu _{2}^{i}(\nu _{3}^{i})^{2}\right] }}{4(\nu
_{2}^{i})^{2}-1}\right] ^{2}\text{ \ otherwise},%
\end{array}%
\right.  \label{GR2D-}
\end{equation}%
Fig. \ref{fig.4} shows the robustness of the mechanical(optical) GR2D $%
\mathcal{D}_{2}^{m}$ ($\mathcal{D}_{2}^{op}$) against the temperature $T$
for various squeezing $r$. The damping ratio $\Gamma $ and the
optomechanical cooperativity $\mathcal{C}$ are the same as in Fig. \ref%
{fig.2}. As expected for quantum discord in a dissipative system, we see
that the GR2D $\mathcal{D}_{2}^{m}$ and $\mathcal{D}_{2}^{op}$ decrease
monotonically when the temperature increases, reaching an asymptotic regime.
In addition, comparing between the results reported in Figs. \ref{fig.2}-\ref%
{fig.4}, we remark that the two quantifiers of nonclassicality $\mathcal{E}%
_{2}^{i}$ and $\mathcal{D}_{2}^{i}$ show rather different behaviors against
the temperature. Indeed, after the total disappearance of the entanglement $%
\mathcal{E}_{2}^{m}$ and $\mathcal{E}_{2}^{op}$, the discord $\mathcal{D}%
_{2}^{m}$ and $\mathcal{D}_{2}^{op}$ still persist and are significantly
non-zero over a wide range of the temperature $T$. We thus conclude that
entanglement is not the only resource comprising quantum discord,i.e.,
two-mode mixed separable states may also contain quantum discord ensuring
the \textit{quantumness of correlations} in such states. More importantly,
Fig. \ref{fig.4} reveals a very special behavior of the Gaussian discord
under thermal noises. In fact, in both mechanical and optical subsystems,
the quantum discord associated with the entangled states has a tendency to
diminish drastically under thermal noises. However the discord associated
with the separable states exhibits a robust behavior,i.e., the asymptotic
value of Gaussian discord may be small but never vanishes remaining almost
constant (\textit{frozen}) even for temperature $T$ up to $0.1~\mathrm{K}$%
(this has been checked for the squeezing values $r=1$ and $r=1.5$). This
behavior reminds us of the freezing phenomenon of quantum correlations
beyond entanglement, where an insightful physical interpretation of such
phenomenon has been discussed in \cite{Kavan Modi,FQC}. Remarkably, the GR2D
$\mathcal{D}_{2}^{m}$ and $\mathcal{D}_{2}^{op}$ are zero only for $r=0.$
This is a consequence of the fact that quantum discord vanishes only if the
Gaussian state is a product state and therefore if and only if the
determinant of the $\mathcal{V}_{3}^{i}$ matrices (see (\ref{CM})) is zero
\cite{Keller2}. Such a condition is well verified for $r=0$, where $c_{\pm
}^{i}=0$ and consequently $\det \mathcal{V}_{3}^{i}=0$ (see Eqs. [(\ref{CM}%
)-(\ref{c})]).\newline
Finally, Fig. \ref{fig.5} shows the GR2D $\mathcal{D}_{2}^{\mathrm{m}}$ and $%
\mathcal{D}_{2}^{op}$ dependence on the optomechanical cooperativity $%
\mathcal{C}$. The mean thermal photons number $n_{\mathrm{th}}$, the damping
ratio $\Gamma $ and the squeezing parameter $r$ are the same as in Fig. \ref%
{fig.3}.
\begin{figure}[t]
\centerline{\includegraphics[width=0.5\columnwidth,height=4.5cm]{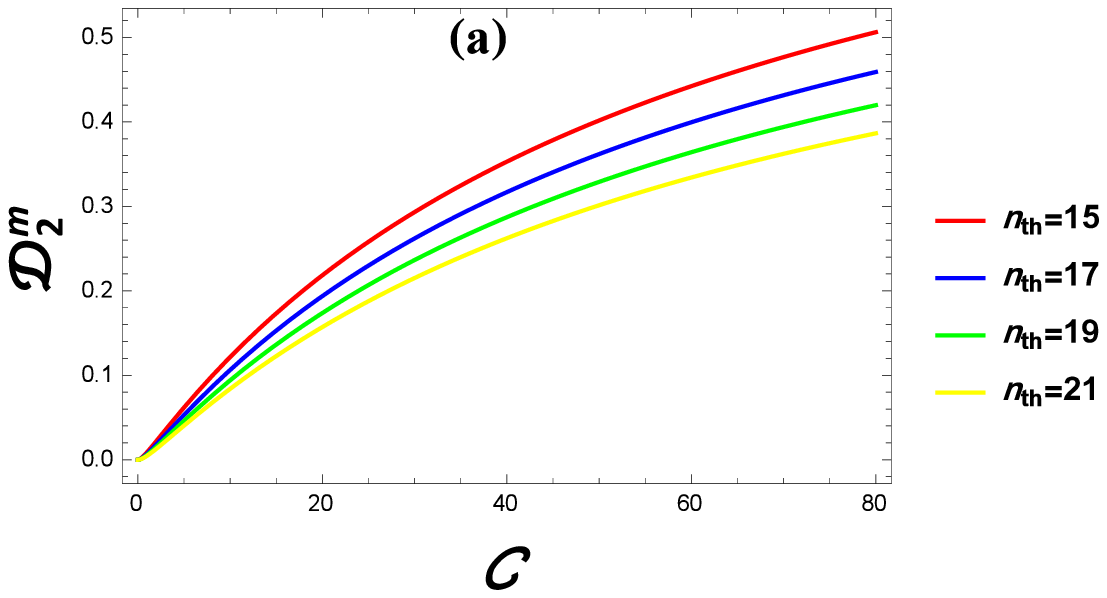}%
\includegraphics[width=0.5\columnwidth,height=4.5cm]{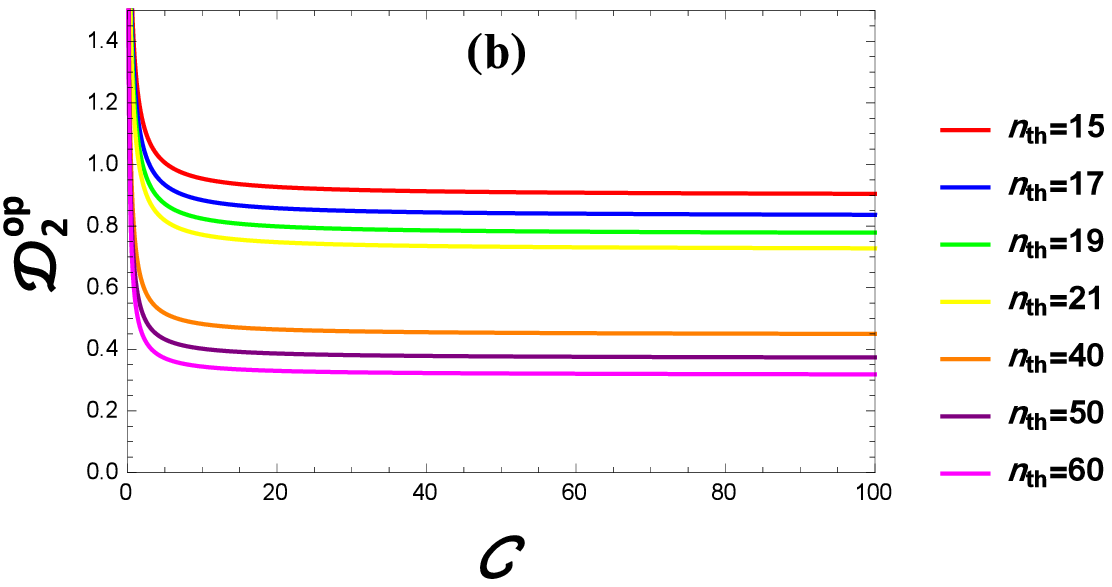}}
\caption{The GR-2 discord $\mathcal{D}_{2}^{i}$ versus the dimensionless
optomechanical cooperativity $\mathcal{C}$ for various values of the thermal
photons number $n_{th}$. (a) Mechanical GR2D $\mathcal{D}_{2}^{m}$, (b)
optical GR2D $\mathcal{D}_{2}^{op}$. The squeezing degree $r$ and the
damping ratio $\Gamma =\frac{\protect\gamma }{\protect\kappa }$ are the same
as in Fig. \protect\ref{fig.3}.}
\label{fig.5}
\end{figure}
Comparing Figs. \ref{fig.3} and \ref{fig.5}, it is clear that when the GR2E $%
\mathcal{E}_{2}^{m}$ and $\mathcal{E}_{2}^{op}$ vanish, the GR2D $\mathcal{D}%
_{2}^{m}$ and $\mathcal{D}_{2}^{op}$ remain non-zero. This result witnesses
the existence of non-classical correlations even in separable states (%
\textit{\ the quantumness of correlations}) of both mechanical and optical
subsystems. \newline
From Fig. \ref{fig.5}, one sees also that like entanglement, quantumness of
correlations can be transferred from light to light or from light to matter.
This transfer is clearly observed from Figs. \ref{fig.4}(a)-\ref{fig.4}(c),
where $\mathcal{D}_{2}^{op}=\mathcal{D}_{2}^{m}=0$ when $r=0$. In contrast $%
\mathcal{D}_{2}^{op}>0$ and $\mathcal{D}_{2}^{m}>0$ for $r\neq 0$, even when
$\mathcal{E}_{2}^{op}$ and $\mathcal{E}_{2}^{m}$ are zero.

\section{Conclusions \label{sec4}}

In two optomechanical Fabry-P\'{e}rot cavities fed by two-mode squeezed
light, the quantumness of correlations of two non-interacting mechanical
modes as well as two non-interacting optical modes have been investigated.
For this, we used respectively the Gaussian R\'{e}nyi-2 entanglement (GR2E)
and the Gaussian R\'{e}nyi-2 discord (GR2D) to quantify bipartite
entanglement and non-classical correlations far beyond entanglement
(quantumness of correlations). The GR2E and GR2D have been analysed under
the same circumstances in terms of the parameters characterizing both the
environment (temperature and squeezing) and the cavities (coupling and
damping). Solving the linearized quantum Langevin equations in the resolved
sideband regime around the classical steady state, we provided analytical
formulas for both mechanical and optical covariances matrices, fully
describing the Gaussian stationary states of the two considered subsystems.
We have seen that in an experimentally accessible parameter regime, that it
is possible to create both quantum entanglement and quantum discord in the
two studied subsystems by quantum fluctuations transfer from either light to
light or light to matter.

The results presented in Sec. \ref{sec3} show clearly that both mechanical
and optical entanglement ($\mathcal{E}_{2}^{m}$ and $\mathcal{E}_{2}^{op}$)
are strongly sensitive to the thermal noises. In particular, the optical
entanglement $\mathcal{E}_{2}^{op}$ is found more robust against thermal
noises than the mechanical entanglement $\mathcal{E}_{2}^{m}$. This makes
optical modes extremely useful in quantum information networks, where they
can be used to transfer states (i.e., information) through dissipative and
noisy channels. In contrast, both mechanical and optical discord $\mathcal{D}%
_{2}^{m}$ and $\mathcal{D}_{2}^{op}$ show a robust behavior under influences
of the thermal baths. Indeed, as shown in Figs. \ref{fig.4}(a)-\ref{fig.4}%
(c), the asymptotic values of the mechanical and optical discord $\mathcal{D}%
_{2}^{m}$ and $\mathcal{D}_{2}^{op}$ may be small but never vanish.
Interesting enough, it has been observed in both mechanical and optical
subsystems that under thermal noises, the discord associated with the
entangled states was found to decay drastically, while the Gaussian discord
measured in the separable states (quantumness of correlations) was
exhibiting a robust behavior. In fact, when the GR2E $\mathcal{E}_{2}^{m}$
and $\mathcal{E}_{2}^{op}$ are non-zero, which corresponds to the entangled
states, the GR2D $\mathcal{D}_{2}^{m}$ and $\mathcal{D}_{2}^{op}$ decrease
aggressively (see Figs. \ref{fig.2} and \ref{fig.4}), whereas when $\mathcal{%
E}_{2}^{m}=\mathcal{E}_{2}^{op}=0$, which corresponds to the separable
states, the GR2D $\mathcal{D}_{2}^{m}$ and $\mathcal{D}_{2}^{op}$ seem to
remain almost constant (frozen) over a wide range of the environmental
temperature (see also Figs. \ref{fig.2} and \ref{fig.4}), meaning that the
quantumness of the correlations can be captured in both mechanical and
optical subsystems even for high temperatures (up to $0.1K$). This has been
checked using squeezing $r=1$ and $r=1.5$.

We notice that, the quantum correlations generated here (entanglement and
quantum discord), could be experimentally verified by homodyning the two
cavities outputs \cite{JieLi}, while the position and momentum of the two
mechanical modes can be measured with the strategy proposed in \cite%
{Vitali(1)}. Finally, the study of stationary evolution of Gaussian R\'{e}%
nyi-2 entanglement and the Gaussian R\'{e}nyi-2 discord may contribute to
the understanding of the quantification of correlations in noisy
environments, which is very interesting for applications in quantum
information processing and communication.

\end{document}